\documentclass[twocolumn,twoside,slac_two, nofootnoteinbib]{revtex4}
\usepackage{graphicx}
\usepackage{fancyhdr}
\usepackage{amsmath}
 \usepackage{amssymb}
 \usepackage{amsfonts}
 \usepackage{eucal}

\newcommand{\be}{\begin{equation}}
\newcommand{\bc}{\begin{center}}
\newcommand{\bse}{\begin{subequations}}
\newcommand{\ese}{\end{subequations}}
\newcommand{\bea}{\begin{eqnarray}}
\newcommand{\eea}{\end{eqnarray}}
\newcommand{\ba}{\begin{array}}
\newcommand{\ea}{\end{array}}
\newcommand{\ee}{\end{equation}}
\newcommand{\ec}{\end{center}}

\def\fillbox#1{\hbox to #1{\vbox to #1{\vfil}\hfil}}

\def\ie{{\it i.e. }}
\def\eg{{\it e.g. }}

\begin{document}


\title{{An $N$-tropic Solution to the Cosmological Constant
Problem}}

%


\author{M. M. Sheikh-Jabbari}
\affiliation{Institute for studies in theoretical Physics and Mathematics (IPM),\\
P.O.Box 19395-5531, Tehran, IRAN}

\begin{abstract}
Based on the assertion that the cosmological constant problem is essentially a
quantum gravity problem, the framework which addresses the cosmological constant problem should
also bear a picture for the ``quantum space-time''. In this talk in an attempt to
address the cosmological constant problem I suggest to start with noncommutative fuzzy spheres
as the toy model for the quantum space-time. In this setting, we show that the cosmological constant
problem may be resolved due to the noncommutativity and ``fuzziness'' of the space and the fact that  the
smallest volume which could be measured in the a quantum space-time is much larger than the naively
expected Planckian size.
This talk is based on \cite{fuzzyholo} which has appeared on the arXiv as hep-th/0605110.
\end{abstract}
\date{\today}
\maketitle

\thispagestyle{fancy}


\section{Introduction}

The Cosmological Constant (CC) problem, has been around since the early days of
conception of the theory of General Relativity: The CC term is the only possibility
which is generally covariant with less than two number of derivatives (of metric), and hence could be added
to the gravity action.  Explicitly, the action
\be\label{EH+Lambda}
 S_{E.H.}={M^2_{Pl}}\int d^4x\ \sqrt{-g} (R+ \frac{2}{M^2_{Pl}}{\Lambda})
\ee
where
\be
M^2_{Pl}\equiv\frac{1}{4\pi G_N},
\ee
describes the Einstein gravity plus the cosmological constant $\Lambda$ whose value is
\emph{ not} determined by any symmetry principle in the ``standard physics'';
its value should be determined through observation, or through a more fundamental theory.
In the above conventions $\Lambda$ is a given constant of dimension of energy density (\ie energy$^4$).

For a long time, based on imprecise cosmological and astrophysical observations, it was believed that
the value of $\Lambda$ should be zero.  The recent cosmic microwave background observations plus the supernovae data,
however, have established  that our Universe is now in an accelerating phase. The simplest, though
not the only, possibility is to attribute this late-time cosmic acceleration to a non-vanishing
CC which has a {\it positive} sign \cite{WMAP3}:
\be\label{CC-value}
{\Lambda}\simeq 3.6\times 10^{-6} GeV\ cm^{-3}= 10^{-120}
M^4_{Pl}.
\ee
As we see the above value is not only non-zero, but also  very small (basically in any physically relevant
scale). This constitutes
what is known as the cosmological constant problem.

To be more precise, there are two essentially different issues which come
under the title of cosmological constant problem.\\
$\S$ The first is a theoretical problem, ``how can one stabilize
the value of the cosmological constant (against quantum
corrections)
in a theoretical setting?''\\
$\S\S$ The second is regarding the value it has according to current observations,
``why $\Lambda$ has the value it has and why it is so small, actually smaller than any other
physical quantity, in the scales natural to a pure gravitational problem,
the Planck mass $M_{Pl}$.
These two problems  sometimes are also  called the ``technical naturalness'' and the ``naturalness'' of the
cosmological constant problem, respectively.

To just show why  both of the CC problems are so non-trivial let us
recall that in a physical model the action  \eqref{EH+Lambda} should be added to the action governing the rest of the
model. From {\it any} quantum field theory viewpoint the CC term, $\sqrt{-\det g} \Lambda$, which is proportional to
{\it identity} operator, is obviously a relevant operator whose coefficient, the cosmological constant
 $\Lambda$, is receiving quantum
corrections and, regardless of the details of the quantum field theory we are using, is UV divergent. If the UV cut-off of the theory is
at $\Lambda_{cutoff}$, then the resulting CC is proportional to the zero point energy which is proportional to
$\Lambda_{cutoff}^4$, the proportionality constant is
positive for bosons and negative for fermions. (The supersymmetric theories use this fact to cancel the zero point
energies, and hence the CC,  by matching the number of bosonic and fermionic degrees of freedom.) On the other hand, from
the gravity theory point of view the CC is an IR problem and has something to do with the large scale structure of the Universe.
Therefore, there is not even a consensus whether the CC problem is a UV or an IR problem.

Here I advocate a different viewpoint that, as we'll see, reconciles the quantum field theory and cosmology viewpoints.
Namely,
 \bc
$\bullet$ {Cosmological Constant is essentially  a Quantum Gravity problem and}\\
$\bullet$ { Quantum Gravity should be formulated on  a
``Quantum Space-time''.}\\
$\bullet$ One of the features of quantum gravity, which is unlike standard quantum
field theories, is presence of  IR/UV mixing. \ec %
The first statement
has been previously alluded to in the literature, \eg see
\cite{Witten} and one may hope that a theory of quantum gravity,
for example string theory, should bear the solution to the CC
problem. Within string theory, however, this has remained a
challenging puzzle. Therefore, here I try to provide an
alternative route utilizing the second statement, starting from
 baby examples of the  quantum space-times and  discuss a
gravity theory which has these ``quantum space-times'' as its
solutions.

In this talk we will address the technical naturalness problem of
the cosmological constant in a {\it Euclidean} setting. Our line
of logic will be as follows: \bc
I) We introduce
the ``quantum'' Euclidean de Sitter space which is a ``quantum''
sphere, the \emph{fuzzy sphere}, $S^4_F$. Fuzzy spheres are described by
$N\times N$ Hermitian matrices and their  radii (in units of Planck length)
is completely determined in terms of the \emph{size of matrices}
$N$.\\
II) We present a \emph{Euclidean} quantum gravity theory which is a
Matrix theory with Matrix-valued vierbein and spin connection as
its degrees freedom.\\
III) We show that in this setting the vacuum
solution is a \emph{fuzzy four sphere}, $S^4_F$. The radius of the vacuum $S^4_F$ solution determines the
(Euclidean) cosmological constant. In our setting, therefore, the
CC is integer-valued and is hence  stable against perturbative quantum corrections.
\ec
As we will argue IR/UV mixing is an inherent feature of the gravity theory we use. Although
it has not been shown explicitly, one expects that if we add matter fields to our matrix gravity
theory, the matrix theory will tame and control the zero point energy contributions.
Note, however, that all the other fields must be also  added in the form of a matrix theory.
This matrix theory is to be constructed
such that in the large matrices (continuum) limit it goes over to the corresponding
standard field theories.

\section{A short introduction to  Fuzzy Spheres}
The idea we use to quantize a given space-time parallels the steps of moving from
classical mechanics to quantum mechanics, where we replace classical phase space by quantum phase space
and Poisson brackets with commutators. We also note that structure of Poisson bracket is  related to the
isometries of the phase space, precisely the volume preserving diffeomorphisms in the phase space.

Let us focus on the case of our interest, the spheres (or
Euclidean de Sitter spaces). First we recall that geometric $S^d$, in an
algebraic setting, is equivalent to the quotient $SO(d+1)/SO(d)$
and everything could be put in the representations of the isometry group $SO(d+1)$.
For the round commutative sphere we are dealing with the \emph{infinite}
dimensional representation
of the $SO(d+1)$. This could be seen, e.g. when we expand
any given function on the $S^d$ in terms of  $SO(d+1)$ spherical harmonics. The latter are generalization of
$Y_{lm}$'s, the $SO(3)$ spherical harmonics. Explicitly, for a given function $\Phi(x)$,
\be\label{Phi-commutative}
\Phi(x)=\sum_{j=0}^{\infty} \Phi_{i_1i_2\cdots i_j}x^{i_1}x^{i_2}\cdots x^{i_j},
\ee
this is reflected in the fact that the sum goes to infinity. In \eqref{Phi-commutative}
$i_k$ is ranging from one to $d+1$ and
\be\label{com-sphere-def}
\sum_{i=1}^{d+1} x_i^2=R^2,
\ee
where $R$ is the radius of the sphere.

The $S^d_F$ is the {\it quantized} or
``fuzzified''  version of  $d$-sphere in such
a way that the $SO(d+1)$ invariance remains intact. This can be
achieved noting the fact that $SO(d+1)$ is a compact group and has
finite dimensional unitary representations. That is, we replace the
continuous coordinates $x_i$ with $N\times N$ matrix coordinates $X_i$, such that
\be\label{Fuzzy-sphere-def}
\sum_{i=1}^{d+1} X_i^2=R^2 {\bf 1}_{N\times N}.
\ee%
Explicit form of the matrices $X^i$ could be worked out using representation theory of
$SO(d+1)$ group, e.g. see \cite{TGMT, SF-soln}, with the generic result that
\be
R^{d-1}\sim N
\ee
for large $N$. Any function (field) on the $S^d_F$ is an $N\times
 N$ matrix. One may expand any field $\Phi(X)$ in terms of
``truncated'' $SO(d+1)$ spherical harmonics:
\be\label{Phi-Fuzzy}
\Phi(X)=\sum_{j=0}^{J_{max}} \Phi_{i_1i_2\cdots i_j} X^{i_1}X^{i_2}\cdots X^{i_j} ,%
\ee%
where $i_{k}=1,2,\cdots, d+1,\ 0\leq k \leq J_{max}$ and
$J_{max}$ is related to the size of matrices $N$ as
\be
N\sim J_{max}^{d-1}.%
\ee%

 Therefore, on the fuzzy spheres there is a \emph{natural,
inherent, cut-off} on the maximum possible angular momentum.

One may define a short length scale on the sphere using the $R$-$N$ relation:%
\be
{l^{d-1}\equiv R^{d-1}/N}
\ee%
or written in a more suggestive way,
\[
{R\sim l J_{max}}\ \ \ {\rm or}\ \ \ {J_{max}\sim X_{max} P_{max}\sim R\times
\frac{1}{l}}
\]
where $X_{max}$ and $P_{max}$ are respectively the maximum value $X$ or momentum $P$ can take on the fuzzy sphere of radius $R$.

Finally, we would like to point out that  the {\emph{ smallest
observable volume}} on $S^d_F$ is \emph{not} $l^d$, but
$L^d$ where
\be\label{smallest-vol}
L^d=(lR)^{d/2}.
\ee
More details on derivation and on discussion of the above can be found in \cite{fuzzyholo}.

\section{The Matrix Quantum Gravity Theory}

We now present a model for quantum gravity, though a Euclidean one,
which has fuzzy spheres among its vacuum solutions.
In this theory the gravitational degrees of freedom, as well as the spacetime coordinates,
are  $N\times N$ hermitian matrices. That is, it is a Matrix \emph{Euclidean}
quantum gravity theory.

This gravity theory, which has been discussed in some detail in
\cite{Nair} and reviewed in \cite{fuzzyholo},
 is based on the Mansouri-Cheng ``gravity as
gauge theory''\cite{F-Mansouri}: In the ordinary (Einstein) gravity
theory which has a group manifold $G/H$ as its vacuum solutions,
the coordinates $x_i,\ i=1,2,\cdots, d\equiv dim G-dim
H$ that parameterize $G/H$ space, are in the (infinite dimensional unitary) representation
of the Lie algebra of $G$,  $g$. In this setting the
gravitational degrees are the ``vierbein''
$e_i^a(x),\ a=1,2,\cdots, d$ and the connection
$\Omega_i^\alpha(x)$, $\alpha=1,2,\cdots, dim H$.\footnote{To be more general and more precise,
the $\alpha$ index is running from one to $dim\ {\rm Env} G-d$, where
${\rm Env} G$ is the enveloping algebra for the fundamental
representation of $g$.} They appear through the covariant
derivative ${\cal D}_i$%
\be%
{\cal D}_i=\partial_i+e_i^a(x) T^a+ \Omega_i^\alpha(x) I^\alpha \ee%
where $I^\alpha\in h$ form a
complete basis for the fundamental representation of the Lie
algebra of $H$, $h$,  $T^a$ the basis for $g-h$, and hence $(T^a,
I^\alpha)$ form a complete basis for $g$.\footnote{In more general cases
the set of $I^\alpha$ should be extended, so that $(T^a,
I^\alpha)$ covers the Enveloping algebra for the fundamental
representation of $g$, ${\rm Env} G$.} As such they are $d\times
d$ unitary matrices. The gravity action is then constructed from
gauge invariant powers (of commutators) of ${\cal D}_i$. In our
example $G=SO(5)$ and $H=SO(4)$ and since the enveloping algebra
of $G$ is other than $G$, it is $U(4)$, $a$ is running from one to
four and $\alpha$ from one to 12. In our example the most natural
from for the action is the Chern-Simons gravity \cite{Nair}.

Using the above setup it is (in a straightforward way) possible to
construct the gravity theory on a noncommutative ``fuzzified'' geometry.
In order that we need to have some knowledge about the representation theory of the
groups $G$ and $H$. In particular, it is important to note that
\emph{if $G$ and $H$ are compact groups},  there exist finite dimensional unitary
$N\times N$ representations. This representation is
naturally embedded in $u(N)$. To formulate our gravity theory we pick this
representation and take the coordinates $x_i$,
$e_i^a(x)$ and $\Omega_i^\alpha(x)$  all to be in this representation. Note that these
are in general  non-commuting. ${\cal
D}_i$ are then taking values in $U(d)\otimes U(N)$. ($U(d)$ is the
enveloping algebra of $G$.) The curvature two-form in the
non-commuting case can again be defined as ${\cal F}_{ij}=[{\cal
D}_i, {\cal D}_j]$.

As the $x_i$'s, and hence the derivatives
$\partial_i$, are non-commuting, ${\cal F}_{ij}$ has a constant
piece. That is this part which leads to the cosmological constant
term in the gravity action.  ${\cal F}_{ij}$ has also a piece
which is proportional to $I^\alpha$. This part contains the
Riemann curvature two form ${\cal R}_{ij}$ and a part proportional
to $T^a$ which is the torsion \cite{Nair}.

In our case, where $d=4$ and $G/H=SO(5)/SO(4)$, we choose $
I^\alpha=\{i\gamma^5,\gamma^a\gamma^5,\gamma^{ab}, i{\bf 1}\}$,
$T^a=i\gamma^a$ and
for the action we take the Chern-Simons action%
 \be\label{action}
S=\kappa \frac{1}{N}{\rm Tr}(\gamma^5\epsilon_{ijkl}{\cal F}_{ij}{\cal F}_{kl})%
\ee%
where the ${\rm Tr}$ is over both the $4\times 4$ and  $N\times N$
matrices. After expanding the above action in terms of the Riemann
curvature and the torsion, what we find is an Einstein-Hilbert
gravity action plus a cosmological constant and some torsional
terms \cite{Nair}. The torsional terms are
proportional to the fuzziness and hence go away in the continuum
limit. The demand that in the continuum (large $N$) limit, and
after proper scaling of the gauge fields and coordinates, we
should recover the usual Einstein gravity, upon assumption
$\ell=l_P$, fixes $\kappa$ as $\kappa^{-1}=R^2 l^2$ which is
equal to the cosmological constant.

The vacuum solutions to the
above gravity theory, by construction, include the fuzzy four
sphere the volume of which and the cosmological constant $\Lambda$, are
related as  $\Lambda^{-1}= R^4N^{-2/3}=L^4 $.
Therefore, the value of the cosmological constant is tied to the
number of degrees of freedom (or the size of the matrices), and being quantized is
protected against perturbative
quantum corrections. That is, in our model we have a way to solve the
technical naturalness of the CC problem.

\section{Discussions and Concluding Remarks}

In this talk we put forward the idea that the solution to the cosmological constant problem
lies in finding a setup in which cosmological constant is quantized and is hence
stable against perturbative (continuous) quantum corrections. We did this, in a Euclidean
setting, by first noting that the radius of a fuzzy sphere is quantized (in Planck units), and
second, introduced a gravity theory which has the fuzzy sphere as its vacuum solution.

One should, however, note that quantization of the CC (in  Planck
units) in itself is not enough to solve the CC problem. For
example, within the string theory setup of flux compactifications
\cite{Bousso-Polch} the value of the four dimensional CC is
proportional to the fluxes and hence quantized. In that case,
unlike ours, there are extremely large number of possibilities
which leads to the string theory ``landscape'' \cite{landscape}.
The CC problem hence re-appears as how/why one of these
possibilities is realized in the real world.

In our model we do not face the above problem as we start with a Matrix gravity theory
with a given size of matrices, $N$. The value of $N$ is among the parameters which
is defining our theory. As we discussed $N$ eventually turns out to be
determining the value of the cosmological constant. In this sense, in our model
the cosmological constant (in Planck units) is a constant of nature (like $\hbar,\ c$
and $l_P$) and is \emph{not} determined dynamically.

Within our approach, however, it is not explicitly seen how the
gravity theory given by \eqref{action} manages to overcome the
usual problem about the contribution of the tadpole diagrams and
zero point energies to the CC. The answer should definitely lie in
the fact that in our gravity theory both UV and IR dynamics of the
gravity are modified due to the noncommutativity which in part
forces us to add some other terms, \eg torsion, to the Einstein
gravity. Exploring this line is postponed to future works.

One of the interesting features of the fuzzy spheres and hence our model, on which we
did not elaborate here, is that the smallest physically observable volume on the fuzzy spheres is not Planckian,
but is much larger ({\it cf.} \eqref{smallest-vol}). One would then expect that
the inverse of the smallest volume $L^{-4}$,
which can also be viewed as the uncertainty
in measurement of the energy density, is related to the vacuum energy density.
As we showed, this is indeed the case in our model.
If we take the radius of the four dimensional fuzzy sphere to be equal to
the Hubble radius today and $l$ to be the Planck length,
we find that $L=\sqrt{lR}\simeq submilimeter$.

Finally,  we would like to stress that here we only discussed the
Euclidean case. Our discussions on the cosmological constant does
not go through for the Minkowski signature, \ie the fuzzy de
Sitter space $dS^4_F$ case, as for this case we should take
$G/H=SO(4,1)/SO(3,1)$ and $SO(4,1)$ is non-compact which has no
finite dimensional unitary representation. The formulation
developed in \cite{Nair-2006} may, however, help to extend our arguments and results
to the more realistic Lorentzian signature.

\section*{Acknowledgements}

I would like to thank the organizers of IPM-LHP06 for organizing the stimulating school and conference.
It is a pleasure to thank Robert Brandenberger, Savas Dimopoulos,
Simeon Hellerman, Esmaeil Mosaffa and  Parameswaran Nair for helpful
comments and especially Eva Silverstein  for  a
suggestion for the title.

\end{document}